\documentclass[aps,prl,twocolumn,showpacs]{revtex4}
\begin{document}
\title{Comment on ``Entanglement Measure for Composite Systems''
 }
\author{Yu Shi}
\email{ys219@phy.cam.ac.uk}
\affiliation{Cavendish Laboratory,
University of Cambridge, Cambridge CB3 0HE, United Kingdom}

\begin{abstract}
A recent paper  claimed to give an entanglement measure for
composite systems, including Bose condensation and
superconductivity. It is not an entanglement measure. It does not
distinguish entanglement and non-factorization merely due to
(anti)symmetrization for identical particles. In fact,  the nature
of entanglement, beyond the effect of (anti)symmetrization, in
important states of many identical particles such as Bose
condensation and superconductivity, had already been studied
earlier.

\end{abstract}

\pacs{03.67.Mn} \maketitle

A recent paper~\cite{yukalov} claimed to give an entanglement
measure for composite systems.  It came to our notice that what
was defined is not an entanglement measure. Moreover, for
identical particles, this measure does not distinguish
entanglement and non-factorization merely due to
(anti)symmetrization, e.g. the Hatree-Fock states, which is not
entangled~\cite{shi2}.

In \cite{yukalov}, a measure $\epsilon(A)$ was defined for an
arbitrary operator $A$ acting on a composite system, which
vanishes when $A$ is a tensor product of operators on subsystems.
Then the operator was taken to be the density matrix $\rho$, and
$\epsilon(\rho)$ was used to measure whether $\rho$ is entangled.
Several examples, including Bose condensation and
superconductivity, were discussed by using this measure.

It can be seen that while $\epsilon(\rho)$ is indeed nonzero for
an entangled pure state of distinguishable particles, it is
nonzero also for an entirely classical mixture.

Moreover, $\epsilon(\rho)$ is nonzero for symmetrized or
antisymmetrized non-entangled state of a system of identical
particles when they occupy more than one single particle states.
In fact, entanglement in a system of identical particles should be
beyond the effect of
(anti)symmetrization~\cite{peres,shi1,zanardi}.

In none of these two cases, the nonzero $\epsilon(\rho)$  measures
the entanglement. This is contrary to the idea in \cite{yukalov}.

Furthermore, the nature of entanglement, {\em beyond the effect of
(anti)symmetrization}, in important states of many identical
particles, such as Bose condensation and superconductivity, had
already been studied in detail earlier~\cite{shi2,shi3}.

\end{document}